\documentclass[twocolumn,
    prc,nofootinbib,tightenlines,superscriptaddress,preprintnumbers]{revtex4-1}

\usepackage[utf8]{inputenc}
\usepackage{newtxtext,newtxmath}
\usepackage{bbold}
\usepackage{bm}

\usepackage{graphicx}
\usepackage[usenames,dvipsnames]{xcolor}
\usepackage{color}
\usepackage{tikz}

\usepackage[colorlinks=true,linkcolor=blue,urlcolor=blue,citecolor=blue]{hyperref}

\usepackage{amsmath,amssymb,amsfonts}
\usepackage{slashed}

\usepackage[english]{babel}
\usepackage{titlesec}
\usepackage{microtype}
\usepackage{dcolumn}

\usepackage{blindtext}

\newcommand{\vect}[1]{\boldsymbol{#1}}

\titlespacing{\section}{4pt}{12pt plus 4pt minus 2pt}{8pt plus 2pt minus 2pt}
\titlespacing{\subsection}{0pt}{12pt plus 4pt minus 2pt}{8pt plus 2pt minus 2pt}
\titlespacing{\subsubsection}{0pt}{12pt plus 4pt minus 2pt}{0pt plus 2pt minus 2pt}

\graphicspath{{.}{figures/}}

\newcommand{\be}{\begin{equation}}
\newcommand{\ee}{\end{equation}}

\def\hs{\hspace}
\def\no{\nonumber}

\begin{document}
\title{Skewness of nuclear matter and three-particle correlations}

\author{Wolfgang Bentz}
\email[]{bentz@keyaki.cc.u-tokai.ac.jp}
\affiliation{Department of Physics, School of Science, Tokai University,
4-1-1 Kitakaname, Hiratsuka-shi, Kanagawa 259-1292, Japan}
%\affiliation{Radiation Laboratory, Nishina Center, RIKEN, Wako, Saitama 351-0198,
%Japan}
\author{Ian C. Clo\"et}
\email[]{icloet@anl.gov}
\affiliation{Physics Division, Argonne National Laboratory, Argonne, Illinois 60439, USA}

\begin{abstract}
We present a study of the skewness of nuclear matter, 
which is proportional to the third derivative of the energy per nucleon with respect to the baryon density at 
the saturation point, in the framework of the Landau-Migdal theory. We derive an exact relation between the 
skewness, the nucleon effective mass, and 
two-particle and three-particle interaction parameters. We also present qualitative estimates,
which indicate that three-particle correlations play an important role for the skewness.   
    \newline\newline
    \noindent \textit{PhySH}: {
        Nuclear matter; 
        Nuclear forces;
        Nuclear many-body theory.
    }
\end{abstract}

\maketitle
%===============================================================================
%===============================================================================
\section{INTRODUCTION}
The properties of infinite nuclear matter have been the subject of
intensive investigations for many decades, mainly because of the wide
range of applications, including the properties of heavy nuclei and their excitation
modes~\cite{Blaizot:1980tw,Mahaux:1985zz}, heavy ion collisions~\cite{Danielewicz:2002pu}, the physics of
supernova explosions~\cite{Bethe:1990mw,Lattimer:2012nd}, and neutron stars~\cite{Baym:2017whm}.
Besides the binding energy per nucleon and the saturation density, other basic properties of 
nuclear matter like the effective mass of nucleons
and its momentum dependence~\cite{Li:2018lpy}, the incompressibility~\cite{Piekarewicz:2009gb}, the symmetry energy and its density dependence~\cite{Zhang:2013wna}, and 
recently the skewness~\cite{Cai:2014kya}, provide basic information on the
effective interaction between nucleons in the medium. Theoretical descriptions of
those and other important properties of nuclear matter have been based on Br\"uckner's 
G-matrix approach~\cite{Bethe:1963rbc,Rajaraman:1963aa,Brown:1971zza}, relativistic theories based 
on meson exchange~\cite{Serot:1984ey,Muller:1995ji,Meng:2016zgn},
effective interactions of the Skyrme~\cite{Zhang:2016aa} and Landau-Migdal type~\cite{Speth:2014tja}, and
effective field theories~\cite{Kaplan:1998we,Hammer:2000xg,Steele:2000qt,Bogner:2009bt,Holt:2017uuq}. 

In this article we focus our attention on the skewness of nuclear matter and its implication
for three-particle correlations. Skewness is defined as $J = 27 \rho^3  \left({\rm d}^3 E_A/{\rm d}\rho^3\right)$ at $\rho=\rho_0$, where $E_A$ is the
energy per nucleon in isospin symmetric nuclear matter, $\rho$ is the baryon density, 
and $\rho_0$ is the saturation density ($\rho_0=0.15$ fm$^{-3}$). This physical quantity has been receiving interest in
connection to phenomenological applications of the liquid drop model to heavy nuclei~\cite{Pearson:1991lsc}, and relativistic approaches to nuclear matter~\cite{Kouno:1995wc,Frohlich:1998zz}.
Recently, empirical values have been extracted in Refs.~\cite{Steiner:2010fz,Cai:2014kya} from flow data
in heavy ion collisions and neutron star observations. Based on the analogy to the
incompressibility, which is defined as $K = 9 \rho^2 \left({\rm d}^2 E_A/{\rm d}\rho^2\right)$ at 
$\rho=\rho_0$, and which is known to be related to the spin-isospin averaged two-particle
interaction in the medium~\cite{Migdal:1967aa,Migdal:1990vm,Kamerdzhiev:2003rd}, one may expect that $J$ is related to the spin-isospin
averaged three-particle interaction. The aim of this paper is to work out this relation
in detail and to discuss the result in connection to empirical values. To achieve this
we will use the framework of the Landau-Migdal theory of nuclear matter~\cite{Migdal:1967aa,Migdal:1990vm}, 
which is based on Landau's Fermi liquid theory~\cite{Landau:1956aa,Landau:1957aa,Landau:1959aa,Baym:2004aa}. The main
advantage of this theory is that symmetries, like gauge invariance and Galilei
invariance (or Lorentz invariance in its relativistic extension~\cite{Baym:1975va})
are incorporated rigorously, and that the description of collective excitations
of the system is physically very appealing~\cite{Nozieres:1964aa,Negele:1988aa}. However,
to the best of our knowledge, the theory has not yet been applied to directly relate the
skewness of nuclear matter to three-particle interactions. It is well known that
the effects of three-particle interactions can be incorporated by using effective
density-dependent two-particle interactions~\cite{Ring:1980aa}, and this approach has been used
in many recent studies with Skyrme-type interactions~\cite{Chen:2009wv}. However, in view of the long history
of studies on three-particle correlations in nuclear matter~\cite{Bethe:1965zz,Day:1981zz,Fritsch:2004nx}, it is
desirable to know a more direct relation of a physical quantity, which is connected 
to observables, to three-particle interaction parameters.   

In Sec.~\ref{sec:II} we derive the relation between the skewness of nuclear matter 
and the three-particle Landau-Migdal parameters, which are defined in a similar way
as the familiar two-particle parameters. In Sec.~\ref{sec:III} we discuss this relation
in connection to empirical values, and in Sec.~\ref{sec:IV} we present further qualitative
discussions, which indicate that three-particle correlations play an important role
for the skewness. A summary is presented in Sec.~\ref{sec:summary},
and additional details are collected in the Appendices. 

%===============================================================================
%===============================================================================
\section{THEORETICAL FRAMEWORK\label{sec:II}}
Here we discuss the form of the first three derivatives of the energy density, $E(\rho)$, of
spin saturated isospin symmetric nuclear matter w.r.t. the baryon density $\rho = 2\,p_{F}^3/(3 \pi^2)$,
where $p_F$ is the Fermi momentum. Extending Landau's formula~\cite{Landau:1956aa,Landau:1957aa,Landau:1959aa,Baym:2004aa} for
the variation of $E$ w.r.t. spin-isospin independent deviations of
the quasiparticle occupation numbers from a reference Fermi distribution (corresponding to 
a density $\rho_0= 2\,p_{F0}^3/(3 \pi^2)>0$) to include the third order term, we can write
\begin{align}
\delta E(\rho) &= 4 \int \frac{{\rm d}^3 k}{(2 \pi)^3}\ \varepsilon(\vect{k}; \rho_0) \, \delta n_{\vect{k}}
\nonumber \\ 
&+ \frac{1}{2}  \left( \prod_{i=1}^2 \, 4 \int \frac{{\rm d}^3 k_i}{(2 \pi)^3} \, 
\delta n_{\vect{k}_i} \right) \, f(\vect{k}_1, \vect{k}_2; \rho_0)  \nonumber \\  
&+ \frac{1}{6}  \left( \prod_{i=1}^3 \, 4 \int \frac{{\rm d}^3 k_i}{(2 \pi)^3} \, 
\delta n_{\vect{k}_i} \right) \, h(\vect{k}_1, \vect{k}_2; \vect{k}_3 ; \rho_0).
\label{vare}
\end{align}
Here $\varepsilon(\vect{k}; \rho)$ is the energy of a quasiparticle with momentum $\vect{k}$;  
$f(\vect{k}_1, \vect{k}_2; \rho)$ is the spin-isospin averaged forward scattering amplitude
of two quasiparticles with momenta $\vect{k}_1, \vect{k}_2$; and
$h(\vect{k}_1, \vect{k}_2; \vect{k}_3 ; \rho)$ is the spin-isospin averaged 
three-particle forward scattering amplitude.
The amplitudes $f$ and $h$ are totally symmetric with respect to interchanges of 
the momentum variables, and 
can be represented by a set of connected diagrams with four and six external nucleon
lines, respectively.
Besides the momentum variables, we also indicate the dependence on the background density explicitly.

The form of $\delta n_{\vect{k}}$ corresponding to a change of the Fermi momentum
($p_{F 0} \rightarrow p_F$) is independent of the direction of $\vect{k}$ and given, 
to first order in $(p_F - p_{F0})$, by
\begin{align}
\delta n_{k} = 
%\theta(p_F - k) - \theta(p_{F0} - k) =
\left(p_F - p_{F 0}\right) \, \delta(p_{F0} - k) \,.   
\label{dn}
\end{align}
The corresponding first order variation of $E$ is given by
\begin{align}
\delta E =  4 \int \frac{{\rm d}^3 k}{(2 \pi)^3} \varepsilon(\vect{k}; \rho_0) \, \delta n_{k}
= \varepsilon(p_{F0}; \rho_0) \,\cdot \, \delta \rho \,,  
\label{var1}
\end{align}
where $\delta \rho = \frac{2 p_{F0}^2}{\pi^2} \cdot (p_F - p_{F0})$.
Because the reference density
$\rho_0>0$ is arbitrary,\footnote{Expansions around $\rho_0=0$ are analytic in $p_F$ but not in $\rho$, and
are not considered in this paper.} we get the well known result~\cite{Hugenholtz:1958zz}  
\begin{align}
\frac{{\rm d} E(\rho)}{{\rm d} \rho} = \varepsilon(p_F; \rho) \,.   
\label{eps}
\end{align}
Next we consider the first order variation of the quasiparticle energy:
\begin{align}
\delta \varepsilon({k}_1; \rho) &=  4 \int \frac{{\rm d}^3 k_2}{(2 \pi)^3} \, 
f(\vect{k}_1, \vect{k}_2; \rho_0) \, \delta n_{{k}_2} \nonumber \\
&= f_0 (k_1, p_{F0}; \rho_0) \, \cdot \, \delta \rho \,,
\label{var2}
\end{align}
where the $\ell=0$ two-particle forward scattering amplitude,
which is proportional to the $s$-wave scattering length in the medium, 
is the angular average of $f({\vect{k}_1}, {\vect{k}_2}; \rho)$.
The definition, together with the $\ell=1$ amplitude which will be used below, is  
\begin{align}
\frac{1}{2 \ell +1} \, f_{\ell} (k_1, k_2; \rho) = 
\int \frac{{\rm d} \Omega_{2}}{4\pi} \,
\left({\vect{\hat k}}_1 \cdot {\vect{\hat k}}_2 \right)^{\ell} \, 
f({\vect{k}_1}, {\vect{k}_2}; \rho),
\label{f01}
\end{align}
where $\ell=0,1$, and we use the notation ${{\rm d} \Omega_i \equiv {\rm d} \Omega_{\vect{k}_i}}$. From Eq.~\eqref{var2} we obtain the partial derivative of the quasiparticle
energy w.r.t. the density as
\begin{align}
\frac{\partial \varepsilon(k_1; \rho)}{\partial \rho} = f_0 (k_1, k_2 = p_{F}; \rho) \,.
\label{deps}
\end{align}
For the total derivative of the Fermi energy (chemical potential)
w.r.t. the density we then obtain the well known result~\cite{Negele:1988aa}\footnote{To simplify the notation, quantities like $f_{\ell}$, $F_{\ell}$ and $M^*$ 
(without arguments) are defined on the Fermi surface. 
A partial derivative $\partial / \partial p_F$ acts only on the momenta, 
$\partial / \partial \rho$  
acts only on the background density, and 
$\left({\rm d}/{\rm d}\rho\right) = \pi^2/(2 p_F^2) \left(\partial/\partial p_F\right) +
\left(\partial / \partial \rho\right)$ is the total derivative w.r.t. the density.}   
\begin{align}
&\frac{{\rm d} \varepsilon(p_F; \rho)}{{\rm d}\rho} =
\frac{{\rm d}^2 E}{{\rm d} \rho^2} = v(p_F; \rho) \, \frac{\pi^2}{2 p_F^2} +
f_0(p_F, p_F; \rho)  \nonumber \\
&\equiv  \frac{\pi^2}{2 p_F M^*(p_F; \rho)} + f_0(p_F, p_F; \rho)
\equiv \frac{\pi^2}{2 p_F M^*} \left(1 + F_0 \right).
\label{second}  
\end{align}
Here $v(k; \rho) = {\rm \partial} \varepsilon(k; \rho)/{\rm \partial} k$ 
is the magnitude of the velocity of the quasiparticle, and we introduced the momentum dependent effective
nucleon mass (Landau effective mass) $M^*(k;\rho)$ by
\begin{align}
v(k; \rho) \equiv \frac{k}{M^*(k;\rho)} \,.
\label{mstar}
\end{align}
In the last relation of Eq.~\eqref{second} we defined the dimensionless Landau-Migdal
parameters as
%usual by~\cite{Migdal:1967aa,Migdal:1990vm,Kamerdzhiev:2003rd}
%
\begin{align}
F_{\ell} = \frac{2 p_F M^*}{\pi^2} \, f_{\ell}  \,.   
\label{dlf} 
\end{align}   
Before proceeding to the third order derivative, we note that the effective mass is related to the 
$\ell=1$ two-particle forward scattering
amplitude by\footnote{For convenience, we collect the relations which arise from
Galilei invariance in App.~\ref{app:A}.} 
\begin{align}
\frac{k}{M^*(k;\rho)} + \frac{2 p_F^2}{3 \pi^2} f_1(k, p_F; \rho) = \frac{k}{M} \,,
\label{gal}
\end{align}
which for $k=p_F$ becomes the familiar Landau effective mass relation~\cite{Landau:1956aa,Landau:1957aa,Landau:1959aa,Nozieres:1964aa}.

Next we consider the derivative of Eq.~\eqref{second} w.r.t. the density, which gives
\begin{align}
\frac{{\rm d}^3 E}{{\rm d} \rho^3} = - \frac{\pi^4}{4 p_F^4 M^*} -
\frac{\pi^2}{2 p_F M^{*2}} \frac{{\rm d} M^*}{{\rm d} \rho} + 
\frac{{\rm d} f_0}{{\rm d}\rho} \,.
\label{h}
\end{align}
By using the derivative of Eq.~\eqref{deps} w.r.t. $k_1$, we can write
\begin{align}
\frac{\partial M^*}{\partial \rho} = 
 - \frac{M^{*2}}{2 p_F} \frac{\partial f_0}{\partial p_F} \,. 
\label{h1}
\end{align}
Recall (see footnote 1) that the quantity $\partial f_{\ell}/\partial p_F$ actually means   
\begin{align}
\frac{\partial f_{\ell}}{\partial p_F} \equiv
\left[\left(\frac{\partial}{\partial k_1} + \frac{\partial}{\partial k_2} \right) 
f_{\ell}(k_1, k_2; \rho)\right]_{k_1=k_2=p_F} \,,  \nonumber  
\end{align}
and because of the symmetry of the 2-particle amplitude this is the same as the derivative 
w.r.t. only one momentum variable, multiplied by $2$. 
Further we note that, from Eq.~\eqref{vare},  
a change of the Fermi momentum leads to the following first order variation 
of the two-particle scattering amplitude:
\begin{align}
\delta f(\vect{k}_1, \vect{k}_2; \rho) &= 4 \int \frac{{\rm d}^3 k_3}{(2\pi)^3} 
\, h(\vect{k}_1, \vect{k}_2, \vect{k}_3; \rho_0) \, \delta n_{{k}_3} \nonumber \\   
&= \delta \rho \,\cdot \, \int \frac{{\rm d} \Omega_{3}}{4 \pi} \, 
h(\vect{k}_1, \vect{k}_2, \vect{k}_3; \rho_0)|_{k_3 = p_{F0}} \,.
\label{df1}
\end{align}
By taking the $\ell=0,1$ moments of this relation, we obtain
\begin{align}
\frac{\partial}{\partial \rho} \, f_{\ell}(k_1, k_2; \rho) =
h_{\ell} (k_1, k_2, k_3=p_F; \rho) \,,    
\label{pf}
\end{align}
where the $\ell=0,1$ three-particle amplitudes are defined by 
\begin{align}
&\frac{1}{2\ell+1} h_{\ell}(k_1, k_2, k_3; \rho)  \nonumber \\ 
&\hs*{10mm}
= \int \frac{{\rm d} \Omega_{2}}{4 \pi} \,\int 
\frac{{\rm d} \Omega_{3}}{4 \pi} \, \left(\vect{\hat k}_1 \cdot \vect{\hat k}_2\right)^{\ell}
h(\vect{k}_1, \vect{k}_2, \vect{k}_3; \rho) \,.
\label{g01} 
\end{align}
Substituting Eqs.~\eqref{h1} and \eqref{pf} into Eq.~\eqref{h} gives
\begin{align}
\frac{{\rm d}^3 E}{{\rm d}\rho^3} = - \frac{\pi^4}{4 p_F^4 M^*} + \frac{3 \pi^2}{4 p_F^2}
\frac{\partial f_0}{\partial p_F} - \frac{\pi^4}{4 p_F^3 M^{*2}} 
\frac{\partial M^*}{\partial p_F} + h_0 \,.
\label{hh}
\end{align}
In order to eliminate the derivative of $f_0$ from this equation, we take the partial derivative
w.r.t. the density on both sides of Eq.~\eqref{gal}, and then set $k=p_F$.
Using again Eq.~\eqref{h1} we obtain
\begin{align}
p_F \,\frac{\partial}{\partial p_F} \left(f_0 + \frac{1}{3} f_1\right) + \frac{4}{3} f_1 
+ \frac{4 p_F^3}{3 \pi^2} \, h_1 = 0 \,.
\label{gal2}
\end{align}
As explained in App.~\ref{app:A}, this relation can also be derived directly from the Galilei invariance
of the two-particle scattering amplitude.
The $f_1$ terms in Eq.~\eqref{gal2} can be expressed by the effective mass, by using 
Eq.~\eqref{gal} and its derivative w.r.t. $k$ at the Fermi surface.
%\begin{align}
%\frac{1}{M^*} - \frac{p_F}{M^{*2}} \frac{\partial M^*}{\partial p_F}
%+ \frac{p_F^2}{3 \pi^2} \frac{\partial f_1}{\partial p_F} = \frac{1}{M}
%\label{a2}
%\end{align}
In this way we can rewrite Eq.~\eqref{gal2} as
\begin{align}
p_F \frac{\partial f_0}{\partial p_F} -
\frac{3 \pi^2}{p_F} \frac{M-M^*}{M M^*} + \frac{\pi^2}{M^{*2}} \frac{\partial M^*}{\partial p_F}
+ \frac{4 p_F^3}{3 \pi^2} \, h_1 = 0 \,.
\label{hhh}
\end{align}
Using this relation to eliminate the derivative of $f_0$ from Eq.~\eqref{hh}
we finally obtain 
\begin{align}
&\frac{{\rm d}^3 E}{{\rm d} \rho^3} =  \frac{3 \pi^4}{4 p_F^4 M^*} \nonumber \\
&\hs*{5mm}
\times \left( -\frac{1}{3}
+ 3 \frac{M - M^*}{M} - \frac{4}{3} \frac{p_F}{M^*} \, \frac{\partial M^*}{\partial p_F}
+ \left(H_0 - H_1\right)  \right) \,.
\label{result}
\end{align}
Here we have defined the dimensionless three-particle interaction parameters by\footnote{As noted in App.~\ref{app:B}, some care has to be taken when comparing the magnitudes of the dimensionless two-particle and three-particle parameters.}
\begin{align}
H_{\ell} &= \left(\frac{ 2 p_F M^*}{\pi^2} \rho\right) \, h_{\ell} = \frac{4 p_F^4 M^*}{3 \pi^4} \, h_{\ell}  \,.     
\label{dlh}
\end{align}
One should note that Eq.~\eqref{result} is an exact relation.

%===============================================================================
%===============================================================================
\section{DISCUSSION ON EMPIRICAL VALUES\label{sec:III}}
In order to connect the relations of Sec.~\ref{sec:II} to empirical values, we note that
the energy per nucleon ($E_A$) is related to the energy density ($E$) by
\begin{align}
E_A(\rho) = E(\rho) /\rho \,.
\label{ea}
\end{align}
All the relations which follow will be given for $\rho = \rho_0$, where $\rho_0$ is the
saturation density defined by 
$\left.{\rm d} E_A / {\rm d} \rho\right|_{\rho = \rho_0}=0$.

Taking the first derivative of Eq.~\eqref{ea} and using Eq.~\eqref{eps} we obtain
the well known relation $E_A(\rho) = \varepsilon(p_F; \rho)$~\cite{Hugenholtz:1958zz}.
The second derivative of Eq.~\eqref{ea} and its relation to the incompressibility ($K$)
of nuclear matter is given by~\cite{Cai:2014kya,Li:2018lpy}
\begin{align}
\frac{{\rm d}^2 E_A}{{\rm d}\rho^2} = \frac{1}{\rho} \frac{{\rm d}^2 E}{{\rm d}\rho^2}
\equiv \frac{K}{9 \rho^2} \,.  
\label{inco}
\end{align}
The third derivative of Eq.~\eqref{ea} and its relation to the skewness ($J$) of nuclear
matter is given by~\cite{Cai:2014kya,Li:2018lpy}
\begin{align}
\frac{{\rm d}^3 E_A}{{\rm d}\rho^3} = - \frac{3}{\rho^2} \frac{{\rm d}^2 E}{{\rm d}\rho^2}
+ \frac{1}{\rho} \frac{{\rm d}^3 E}{{\rm d}\rho^3} \equiv  \frac{J}{27 \rho^3} \,.  
\label{skew}
\end{align}
By using Eqs.~\eqref{second} and \eqref{result}, the incompressibility and the skewness can
be expressed as 
\begin{align}
K &= \frac{3 p_F^2}{M^*} \, \left(1 + F_0 \right)  \,,
\label{k} \\
J &= - 9 K + \frac{9 p_F^2}{M^*}  \nonumber \\
&\times \left(-\frac{1}{3} + 3 \frac{M-M^*}{M} - \frac{4}{3} \, \frac{p_F}{M^*} \, \frac{\partial M^*}{\partial p_F}
+ \left(H_0-H_1\right) \right) \,.
\label{j}
\end{align}
%
%All quantities in the above relations, except for the
%three-particle terms in Eq.~\eqref{j}, have been the subject of intensive investigations.
Recent analyses~\cite{Cai:2014kya,Li:2018lpy} which take into account constraints from heavy ion collisions
and astrophysical observations, suggest the following ranges 
of $K$ and $J$ at the saturation density $\rho_0 = 0.15$ fm$^{-3}$:
\begin{align}
0.20 \, [{\rm GeV}]  \, < \,\,  &K \, < \,\, 0.30  \, [{\rm GeV}],    
\label{kemp} \\    
-0.50 \, [{\rm GeV}] \, < \,\,  &J  \, < \,\, -0.01 \, [{\rm GeV}].
\label{jemp}
\end{align}
From these values and Eqs.~\eqref{k} and \eqref{j} we can make a few observations:
First, the Fermi gas value of the incompressibility $K_{\rm free} = 3 \frac{p_F^2}{M} \simeq 0.22$ 
GeV is inside the range of Eq.~\eqref{kemp}, i.e., the detailed values of $M^*$ and $F_0$ result in fine-tuning
of $K$ but do not lead to order of magnitude changes. On the other hand, the Fermi gas value
for the skewness $J_{\rm free}= - 10 K_{\rm free} \simeq -2.2$ GeV is very large in magnitude compared to the
empirical values of Eq.~\eqref{jemp}, indicating that some of the correction terms in
Eq.~\eqref{j} must be large and positive. In fact, by using the lower limits of Eqs.~\eqref{kemp} and \eqref{jemp}, we see that Eq.~\eqref{j} implies 
\begin{align}
\frac{8}{3} \, \frac{M-M^*}{M^*} - \frac{4}{3} \frac{p_F M}{M^{*2}} \frac{\partial M^*}{\partial p_F}
+ \frac{M}{M^*}\left(H_0-H_1\right) > 2.38.
\label{ine}
\end{align}
The value of the effective mass at the saturation density of symmetric nuclear matter has been the subject of
intensive investigations during the last decades~\cite{Mahaux:1985zz,Li:2018lpy}, and most of them are consistent with the range
\begin{align}
0.7 \, < \, \frac{M^*}{M} \, < \,\, 1.0.  \label{msemp}
\end{align}  
The momentum dependence of the effective mass has been intensively studied by Mahaux {\it et al.}~\cite{Mahaux:1985zz}.
Their studies, as well as subsequent work of Blaizot {\it et al.}~\cite{Blaizot:1981zz}, indicate that $M^*(k;\rho)$
has a pronounced peak very close to the Fermi surface 
(i.e., $\partial M^*/\partial p_F \simeq 0$), and it was shown
that this behavior comes mainly from the same polarization effects which cause an imaginary part 
of the single particle energy away 
from the Fermi surface. More recent studies~\cite{vanDalen:2005sk} indicate
that this peak of the effective mass may be slightly above the Fermi surface, i.e., $M^*(k;\rho)$ is still 
increasing at $k=p_F$ before it reaches a maximum.\footnote{Fig.~4 of Ref.~\cite{vanDalen:2005sk} indicates that
$\frac{{\rm \partial} M^*}{{\rm \partial} p_F} \simeq 1.8$ at the saturation density.} In our subsequent estimates we will
therefore assume that $\partial M^*/\partial p_F > 0$. We then obtain an upper limit for the first two terms in Eq.~\eqref{ine}:
\begin{align}
\frac{8}{3} \, \frac{M-M^*}{M^*} - \frac{4}{3} \frac{p_F M}{M^{*2}} \frac{\partial M^*}{\partial p_F}
< 1.14. 
\label{ine1}
\end{align}
By comparing Eqs.~\eqref{ine} and \eqref{ine1} we clearly see that one needs a large positive contribution
from the three-body terms in order to explain the empirical values:
\begin{align}
\frac{M}{M^*} \left(H_0 - H_1 \right) > 1.24.  
\label{in}
\end{align}

%===============================================================================
%===============================================================================
\section{QUALITATIVE DISCUSSIONS ON THE THREE-PARTICLE AMPLITUDE\label{sec:IV}}
%\setcounter{equation}{0}
%
%===============================================================================
\begin{figure}
\includegraphics[scale=0.4]{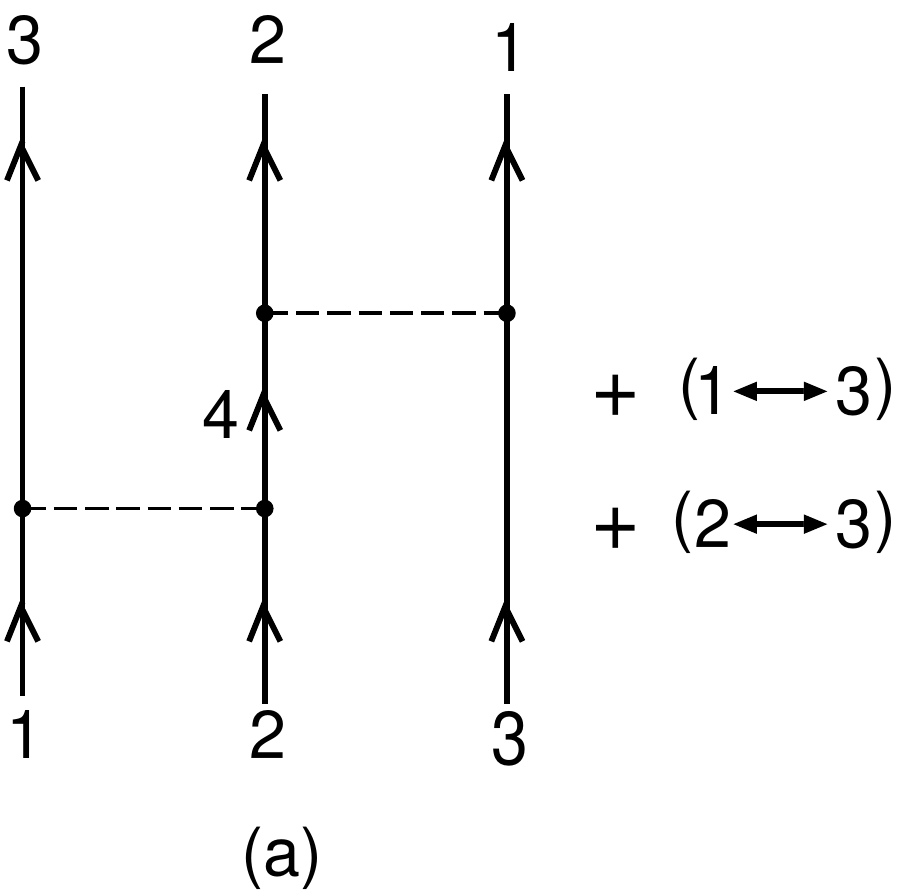}
\hspace{1cm} 
\includegraphics[scale=0.4]{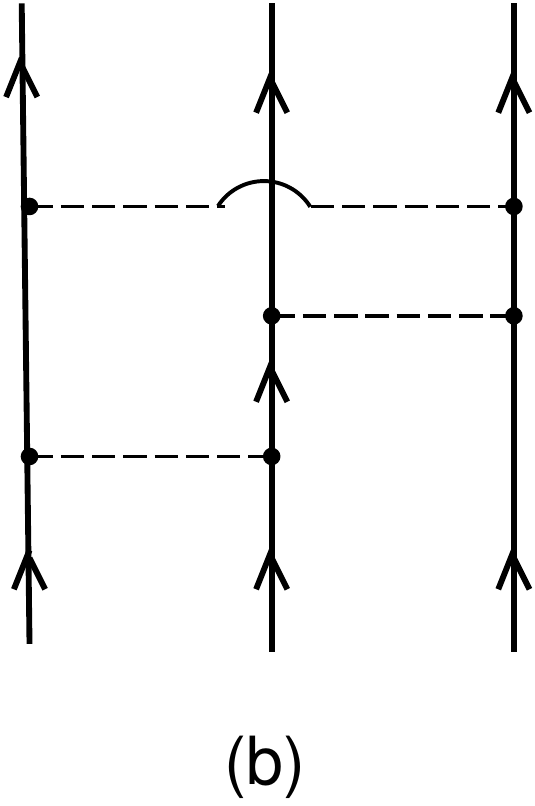}
\caption{(a) Lowest order diagram contributing to the two-particle correlation piece $h^{\rm (2pc)}$.
(b) An example of a contribution to the three-particle correlation piece $h^{\rm (3pc)}$. 
In each case the solid lines are nucleons, and the dashed lines denote the nuclear potential, which is assumed
to be antisymmetrized according to Eq.~\eqref{va}.}
\label{fig:1}
\end{figure}
%===============================================================================

The three-particle amplitude $h(\vect{k}_1,\vect{k}_2,\vect{k}_3;\rho)$ contains 1-particle reducible
pieces, i.e., diagrams which can be made disconnected by cutting a single intermediate 
nucleon line, and 1-particle irreducible pieces.
In perturbation theory, the former ones start with terms of second order in the nuclear potential,
see Fig.~\ref{fig:1}\textcolor{blue}{(a)}. We will refer to them as the 
``two-particle correlation'' (2pc) contributions to the three-particle amplitude.   
On the other hand, the 1-particle irreducible pieces start with third order perturbation theory, and
involve interactions between all three pairs of the particles. (See Fig.~\ref{fig:1}\textcolor{blue}{(b)} for an example.)
We will call them the ``three-particle correlations'' (3pc). We therefore split the three-particle amplitude as follows:      
\begin{align}
h = h^{(\rm 2pc)} + h^{(\rm 3pc)}  \,.    
\label{sum}
\end{align}
\looseness=-1
The expression for the lowest order process shown in Fig.~\ref{fig:1}\textcolor{blue}{(a)} can be
derived from the definition given in Eq.~\eqref{vare} by taking the second functional derivative of the 
energy density in second order perturbation theory. 
Those familiar relations~\cite{Brown:1971zza,Negele:1988aa} are summarized in App.~\ref{app:B}, and 
the second (2nd) order result for $h$ is 
\begin{align}
&h^{(\rm 2nd)} (\vect{k}_1, \vect{k}_2, \vect{k}_3; \rho) =
\sum_4 \delta_{\vect{k}_1 + \vect{k}_2, \vect{k}_ 3 + \vect{k}_4}   \nonumber \\
&\hs*{13mm}
\times P \, \frac{|\langle 12| \, \hat{v} \, | 34 \rangle_a|^2}{ \epsilon_3 + \epsilon_4 - \epsilon_1 - \epsilon_2}  
+ \left( 1 \leftrightarrow 3 \right) +  \left( 2 \leftrightarrow 3 \right) \,.  
\label{expr}
\end{align}
In this schematic notation, $1\sim 4$ represent the momenta $\vect{k}_1 \sim \vect{k}_4$ as well as
the associated spin and isospin components, though an average over the spin and isospin components 
of $1$, $2$, $3$ is assumed implicitly. 
The sum represents momentum integration and summation over spin and isospin components, the $\delta$ symbol 
represents a momentum conserving $\delta$-function, $P$ denotes
the principal value,
and the $\epsilon$'s represent the free nucleon kinetic energies $\epsilon_i = k_i^2/(2M)$. 
The antisymmetrized matrix elements of the nuclear potential ($\hat{v}$) are defined by
\begin{align}
\langle 12| \, \hat{v} \, |34 \rangle_a = \langle 12| \, \hat{v} \, | 34 \rangle
- \langle 12| \, \hat{v}  \, |43 \rangle \,.
\label{va}
\end{align}
The fact that there is no Fermi step-function $n_4$ associated with the intermediate nucleon 4 
in Eq.~\eqref{expr} is consistent with the 1-particle reducible character: 
If there were a dependence on $n_4$, one further variation could be carried out to
derive a connected four-particle amplitude. This, however, is not possible for the
process shown in Fig.~\ref{fig:1}\textcolor{blue}{(a)}, because cutting the intermediate nucleon line gives
a disconnected diagram.

%===============================================================================
\begin{figure}
\centering
\includegraphics[scale=0.4]{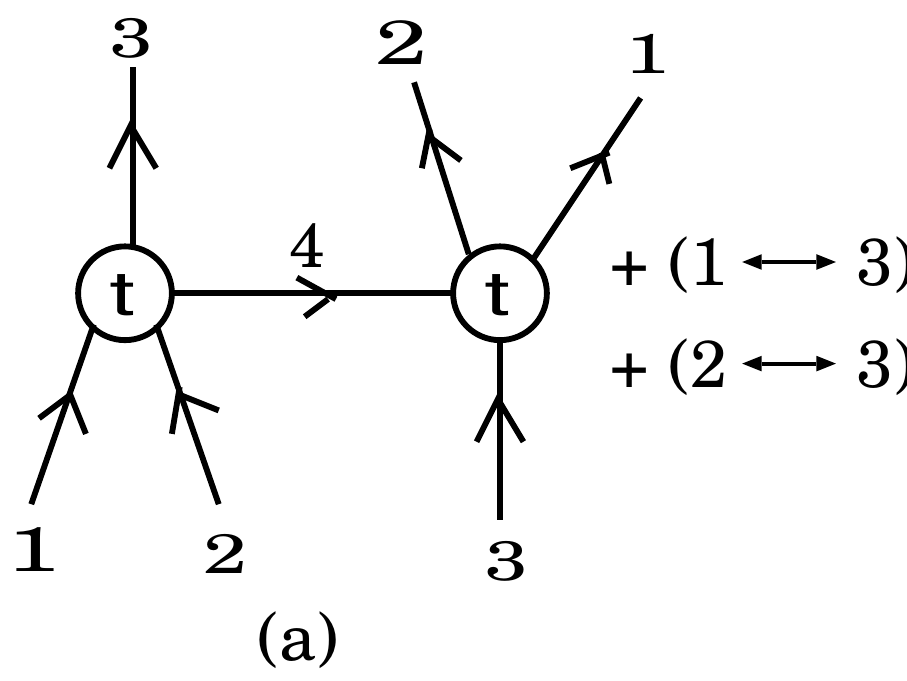}
\hspace{1cm} 
\includegraphics[scale=0.4]{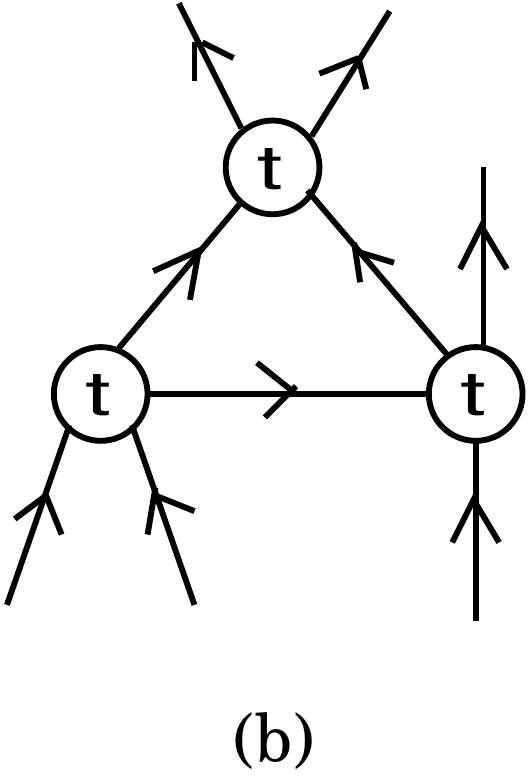}
\caption{(a) The two-particle correlation piece $h^{\rm (2pc)}$ of Eq.~\eqref{expr1}. (b) An example of
the three-particle correlation contribution $h^{\rm (3pc)}$. In each case the solid lines are nucleons, 
and $t$ represents a 2-body off-forward scattering matrix. The two diagrams actually represent
the first two terms in the Faddeev series.}
\label{fig:2}
\end{figure}
%===============================================================================

A few points of this simple perturbative example can be generalized 
(see also App.~\ref{app:B} for examples):
According to the definition given in Eq.~\eqref{vare}, the function $h$
can be obtained from a given diagram for the energy density (effective
potential) by successively cutting three intermediate nucleon lines which we denote by 
1, 2, and 3.\footnote{We can draw the resulting diagram for $h$ in the ``particle channel'',
i.e., if a cut line ($p$) was {\em beginning} at an interaction vertex $A$ and {\em terminating}
at a vertex $B$, we draw a particle line $p$ {\em outgoing} from $A$ and {\em incoming} into $B$.} 
If the cuts are taken in such a
way that, among the remaining lines, there is one (we denote it by 4) 
whose momentum is fixed by momentum conservation, then the four lines 1, 2, 3, 4  
constitute the external legs of a two-particle scattering
amplitude (matrix elements of an operator $\hat{t}$), which is the 
off-forward generalization of the function $f$ defined by
Eq.~\eqref{vare}. In this case, the resulting diagram can be made disconnected by
cutting the line 4, i.e., this line must connect two 2-particle irreducible $t$-blocks, 
as shown in Fig.~\ref{fig:2}\textcolor{blue}{(a)}. The expression for this process is then obtained by a generalization of 
Eq.~\eqref{expr}, i.e., replacing $\hat{v} \rightarrow \hat{t}$ and $\epsilon_i \rightarrow \varepsilon_i$, 
where the $\varepsilon_i$'s are the quasiparticle energies:
\begin{align}
&h^{(\rm 2pc)}(\vect{k}_1, \vect{k}_2, \vect{k}_3; \rho) =
\sum_4 \delta_{\vect{k}_1 + \vect{k}_2, \vect{k}_ 3 + \vect{k}_4} \nonumber \\
&\hs*{12mm}
\times P \, \frac{|\langle 12| \, \hat{t} \, | 34 \rangle_a|^2}
{ \varepsilon_3 + \varepsilon_4 - \varepsilon_1 - \varepsilon_2}  
+ \left( 1 \leftrightarrow 3 \right) +  \left( 2 \leftrightarrow 3 \right). 
\label{expr1}
\end{align}
The energy denominator in this expression comes 
from the intermediate state where only the particles 1, 2, 3, 4 are
``in the air'', and is the same irrespective of whether particle 4 propagates forward 
or backward in time. Hence, there is no dependence on the Fermi step function $n_4$ of particle 4,
which is consistent with the 1-particle reducible character of the process, as explained
above for the lowest order case. On the other hand, if the cuts in the diagram for the energy density are taken
in any other way, the resulting diagram for $h$ is 1-particle irreducible, and the
example shown in Fig.~\ref{fig:2}\textcolor{blue}{(b)} is a generalization of the lowest order process of Fig.~\ref{fig:1}\textcolor{blue}{(b)}. 
We see that actually the process of Fig.~\ref{fig:2}\textcolor{blue}{(a)} is the driving term of the Faddeev
equation (one particle is exchanged between an interacting pair and the third particle),
and Fig.~\ref{fig:2}\textcolor{blue}{(b)} appears in the first iteration of the Faddeev equation~\cite{Bethe:1965zz,Day:1981zz,Fritsch:2004nx}. 

In order to get a rough estimate of the two-particle correlation piece $h^{\rm (2pc)}$ and its
contribution to the inequality given in Eq.~\eqref{in},
we assume that the 2-particle $t$-matrix in Eq.~\eqref{expr1} can be represented by an effective contact interaction,
i.e., by the in-medium scattering length~\cite{Fetter:1971aa}. 
In this case, the angular averages of Eq.~\eqref{g01} concern only the energy denominator of 
Eq.~\eqref{expr1}, and with the further assumption that the
latter can be obtained from the free expression in Eq.~\eqref{expr} by the replacement
$M \rightarrow M^*$, assuming an average value of the effective mass in the 
range given in Eq.~\eqref{msemp}, the angular integrals can be done analytically (see App.~\ref{app:C}):
\begin{align}
&\int \frac{{\rm d} \Omega_{2}}{4 \pi} \,\int \frac{{\rm d} \Omega_{3}}{4 \pi}
\left(\frac{P}{ \varepsilon_3 + \varepsilon_4
-\varepsilon_1 -\varepsilon_2}  + (1 \leftrightarrow 3) + (2 \leftrightarrow 3) \right) \nonumber \\ 
&= \frac{M^*}{p_F^2} \, \left( 3 \ln 2 \right) \,, 
\label{int1} 
\end{align} 
\begin{align}
&3 \, \int \frac{{\rm d} \Omega_{2}}{4 \pi} \,\int \frac{{\rm d} \Omega_{3}}{4 \pi} \,
\left(\vect{\hat k}_1 \cdot \vect{\hat k}_2\right)  \nonumber \\
&\hs*{20mm}
\times \left( \frac{P}{ \varepsilon_3 + 
\varepsilon_4 -\varepsilon_1 -\varepsilon_2 } + (1 \leftrightarrow 3)
+ (2 \leftrightarrow 3) \right) \nonumber  \\
&= - \frac{M^*}{p_F^2} \, \left(1 - \ln 2 \right)  \,. 
\label{int2}
\end{align}
Because these simple expressions indicate that $\ell=1$ contributions are suppressed by large factors
compared to the $\ell=0$ contributions, we neglect the $\ell=1$ terms in the following.
To be specific, we assume that the matrix elements $_a\langle 34| \,\hat{t}\, |12\rangle$ 
can be replaced by the $\ell=0$ part of an effective interaction of the Landau-Migdal type~\cite{Migdal:1967aa,Migdal:1990vm,Kamerdzhiev:2003rd}:
\begin{align}
&_a\langle 34 | \, \hat{t} \,|12 \rangle = f_0 \, \left(\delta_{31} \cdot \delta_{42}\right) + f_0' \, 
\left(\vect{\tau}_{31} \cdot \vect{\tau}_{42} \right) \nonumber \\ 
&\hs*{12mm}
+ g_0 \, \left(\vect{\sigma}_{31} \cdot \vect{\sigma}_{42} \right)
+ g_0' \, \left(\vect{\sigma}_{31} \cdot \vect{\sigma}_{42} \right) \, 
\left(\vect{\tau}_{31} \cdot \vect{\tau}_{42} \right),
\label{lm}
\end{align}
where the notation indicates that the spin and isospin operators are defined to act in the particle-hole channel.
As usual, the effect of exchange terms is assumed to be included in the interaction parameters. 
Performing then the spin-isospin sum over 4 as well as the spin-isospin averages over 1,\,2,\,3 in Eq.~\eqref{expr1}, we obtain
\begin{align}
h_0^{\rm (2pc)} = \frac{M^*}{p_F^2} \cdot \left(3 \ln 2\right) \cdot
\frac{1}{4} \left(f_0^2 + 3 f_0^{'2} + 3 g_0^2 + 9 g_0^{'2} \right) \,.
\label{final1}
\end{align}
In terms of the dimensionless interaction parameters of Eqs.~\eqref{dlf} and \eqref{dlh}, this can be expressed as
\begin{align}
H_0^{\rm (2pc)} = \ln 2  \cdot 
\frac{1}{4} \left(F_0^2 + 3 F_0^{'2} + 3 G_0^2 + 9 G_0^{'2} \right).
\label{final2}
\end{align}
In order to estimate the 2pc contributions to the inequality given in Eq.~\eqref{in}, we refer to recent
calculations done by using several parameter sets of the extended Skyrme interaction~\cite{Zhang:2016aa} 
and chiral effective
field theory~\cite{Holt:2017uuq}. (Both calculations are consistent with the ranges given in Eq.~\eqref{kemp} 
for the incompressibility and Eq.~\eqref{msemp} for the effective mass.) Using the values for the Landau-Migdal parameters 
given in Table~IV of Ref.~\cite{Zhang:2016aa}, and in Figs.~9 and 10 of Ref.~\cite{Holt:2017uuq} at the saturation density, 
we see that the values for
the quantity $\frac{M}{M^*}\,H_0^{\rm (2pc)}$ are between $0.75$ and $1.23$.
Comparison with Eq.~\eqref{in} then indicates that the 2pc piece alone 
is too small to reproduce the empirical results. This gives us a hint that
the three-particle correlation term in Eq.~\eqref{sum} plays an important role
for the skewness of nuclear matter. 
%More detailed investigations, using for example the Faddeev method in the framework of
%effective field theories~\cite{Hammer:2000xg,Steele:2000qt,Bogner:2009bt,Hammer2000}, are necessary to support this observation.

%===============================================================================
%===============================================================================
\section{SUMMARY\label{sec:summary}}
In this paper we used the framework of the Landau-Migdal theory to express
the skewness of nuclear matter in terms of the effective nucleon mass, its slope as a function
of momentum, and two-body as well as three-body interaction parameters.
Our main result is summarized by the formula given in Eq.~\eqref{j}, where $H_0$ and $H_1$ are the
dimensionless three-body Landau-Migdal parameters, which we defined 
in Eq.~\eqref{dlh} via the $\ell=0$ and $\ell=1$ moments of the three-body
forward scattering amplitude $h$. We pointed out that, in order to explain
the range of empirical values for $J$ given in Eq.~\eqref{jemp}, one needs
a large positive contribution from the three-body interaction terms, as
expressed by Eq.~\eqref{in}. We attempted to make a
rough estimate of the contributions of two-nucleon correlations 
[see Figs.~\ref{fig:1}\textcolor{blue}{(a)} and \ref{fig:2}\textcolor{blue}{(a)}] to those three-body parameters, and found that
they are of the right sign but very likely too small to explain the
empirical values of $J$. We find this to be an important hint that
three-particle correlation processes, like those shown in Figs.~\ref{fig:1}\textcolor{blue}{(b)} and
\ref{fig:2}\textcolor{blue}{(b)}, play an important role for the skewness of nuclear matter.

Usually the effect of three-particle interactions are incorporated
by using effective density-dependent two-particle interactions. 
In view of the long and important history of studies on three-particle correlations
in nuclear matter, however, it would be interesting to assess our
results more quantitatively, by using for example the 
Faddeev method in the framework of effective field theories.

%===============================================================================
%===============================================================================
\begin{acknowledgments}
W.B. acknowledges the hospitality of Argonne National Laboratory, where part of this work has been performed, and expresses his thanks to Prof. J. Speth and Prof. H. Sagawa for their helpful correspondence. The work of I.C. was supported by the U.S. Department of Energy, Office of Science, Office of Nuclear Physics, contract no. DE-AC02-06CH11357.
\end{acknowledgments}

%===============================================================================
%===============================================================================
\appendix
\section{GALILEI INVARIANCE\label{app:A}}
For completeness we review here the derivation of the Landau relation between $M^*$ and $f_1$: 
One considers the variation of the quasiparticle energy
which arises from the change of the distribution function due to a Galilei transformation
from the rest system of nuclear matter to a system which moves with velocity
$\vect{u} \equiv \vect{p}/M$, where $M$ is the free nucleon mass. 
To first order in $p$ these variations are given by 
\begin{align}
\delta n_{\vect{k}} &= 
%\theta \left(p_F - |\vect{k} + \vect{p}| \right)  - \theta(p_F - k)
 - (\vect{\hat k} \cdot \vect{p}) \, \delta(p_F - k) \,,
\label{gal1} \\
\delta \varepsilon(\vect{k}; \rho) &=  4 \int \frac{{\rm d}^3 k_2}{(2 \pi)^3} \, 
f(\vect{k}, \vect{k}_2; \rho) \, \delta n_{\vect{k}_2}   \nonumber \\
&= - \frac{2 p_F^2}{3 \pi^2} \left(\vect{\hat k} \cdot \vect{p}\right) \, f_1(k, p_F; \rho) \,.
\label{gala} 
\end{align}
On the other hand, the quasiparticle energy should transform in the same way as a Hamiltonian
in classical mechanics, i.e;
$\varepsilon'(\vect{k}';\rho) = \varepsilon(\vect{k};\rho) - \frac{\vect{k}\cdot\vect{p}}{M}
+ \frac{p^2}{2M}$, where $\vect{k}' = \vect{k} - \vect{p}$. From this it follows that
$\varepsilon'(\vect{k};\rho) = \varepsilon(\vect{k} + \vect{p}; \rho) - 
\frac{\vect{k}\cdot \vect{p}}{M} - \frac{p^2}{2M}$, and to first order in $p$, 
\begin{align}
\delta \varepsilon(\vect{k}; \rho)   
= \left(\vect{k} \cdot \vect{p} \right) \left(\frac{1}{M^*(k,\rho)} - \frac{1}{M} \right)\,,
\label{varg}
\end{align}
where we used the definition given in Eq.~\eqref{mstar}.
The requirement that Eqs.~\eqref{gala} and \eqref{varg} are identical leads to the desired relation
of Eq.~\eqref{gal} in the main text:
\begin{align}
\frac{k}{M^*(k;\rho)} + \frac{2 p_F^2}{3 \pi^2} f_1(k, p_F; \rho) = \frac{k}{M} \,,
\label{gal0}
\end{align}
which  holds for any values of $k$ and $\rho$. For the case $k=p_F$, this becomes the familiar 
Landau effective mass relation.

As explained in the main text, by taking derivatives of Eqs.~\eqref{deps} and \eqref{gal} one can derive Eq.~\eqref{gal2}. Here we show that this relation can also be derived by the
requirement of Galilei invariance of the 2-particle forward scattering amplitude:
Under the change given by Eq.~\eqref{gal1} of the quasiparticle distribution function, the scattering amplitude
changes according to 
\begin{align}
&\delta f(\vect{k}_1, \vect{k}_2; \rho) = - 4 \int \frac{{\rm d}^3 k_3}{(2\pi)^3} \no \\
&\hs*{24mm}
\, h(\vect{k}_1, \vect{k}_2, \vect{k}_3; \rho)
\left(\vect{p} \cdot \vect{\hat k}_3\right) 
\delta(p_F - k_3) \,. 
\label{transf}
\end{align}
On the other hand, the Galilei invariance of the scattering amplitude is expressed as
$f'(\vect{k}'_1, \vect{k}'_2; \rho) = f(\vect{k}_1, \vect{k}_2; \rho)$, where
$\vect{k}_i'= \vect{k}_i - \vect{p}$. From this it follows, to first order in $p$, that  
$\delta f(\vect{k}_1, \vect{k}_2; \rho) = \vect{p} \cdot 
\left({\boldsymbol \nabla}_{k_1} + {\boldsymbol \nabla}_{k_2}\right) \, f(\vect{k}_1, \vect{k}_2; \rho)$.
The requirement that this change is the same as Eq.~\eqref{transf} gives
\begin{align}
&\left({\boldsymbol \nabla}_{k_1} + {\boldsymbol \nabla}_{k_2}\right) \, 
f(\vect{k}_1, \vect{k}_2; \rho) \no \\
&\hs*{10mm}
= - 4 \int \frac{{\rm d}^3 k_3}{(2\pi)^3} 
\, h(\vect{k}_1, \vect{k}_2, \vect{k}_3; \rho) \, \vect{\hat k}_3  \, \delta(p_F - k_3)\,.
\end{align}
Multiplying this relation by $\vect{\hat k}_1$  and integrating over the directions of
$\vect{k}_2$ this becomes
\begin{align}
&\int \frac{{\rm d}\Omega_2}{4\pi} \left(\vect{\hat k}_1 \cdot {\boldsymbol \nabla}_{k_1}
+ \vect{\hat k}_1 \cdot {\boldsymbol \nabla}_{k_2} \right) f(\vect{k}_1, \vect{k}_2; \rho)
\nonumber \\
&= - \frac{2 p_F^2}{\pi^2} \int \frac{{\rm d}\Omega_2}{4\pi}  
\int \frac{{\rm d}\Omega_3}{4\pi} \left(\vect{\hat k}_1 \cdot \vect{\hat k}_3 \right) 
h(\vect{k}_1, \vect{k}_2, \vect{k}_3; \rho)|_{k_3=p_F} \,.
\nonumber \\
\label{s}
\end{align} 
We consider this relation for the case $k_1 = k_2 = p_F$.  
In the first term on the l.h.s. we simply have $\vect{\hat k}_1 \cdot {\boldsymbol \nabla}_{k_1}
= \partial / \partial k_1$. For the second term, we can use a partial integration in
$\vect{k}_2$ to show that
\begin{align}
&\int \frac{{\rm d}\Omega_2}{4\pi}  \left(\vect{\hat k}_1 \cdot {\boldsymbol \nabla}_{k_2} \right) 
f(\vect{k}_1, \vect{k}_2; \rho)  \nonumber \\
&= \int \frac{{\rm d}\Omega_2}{4\pi} \left(\hat{\vect{k}_1} \cdot \hat{\vect{k}_2} \right)
\left(\frac{2}{p_F} + \frac{\partial}{\partial k_2} \right) f(\vect{k}_1, \vect{k}_2; \rho)
\nonumber \\
&= \frac{1}{3} \left(\frac{2}{p_F} + \frac{\partial}{\partial k_2}\right) f_1(k_1, k_2; \rho).
\nonumber
\end{align}
Substituting this for the second term on the l.h.s. of Eq.~\eqref{s}, and using the definition 
given in Eq.~\eqref{g01} as well as the symmetry of $h$ in the momentum variables, leads to Eq.~\eqref{gal2}.

%===============================================================================
%===============================================================================
\section{PERTURBATION THEORY\label{app:B}}
\begin{figure}[tbp]
\centering
\includegraphics[scale=0.4]{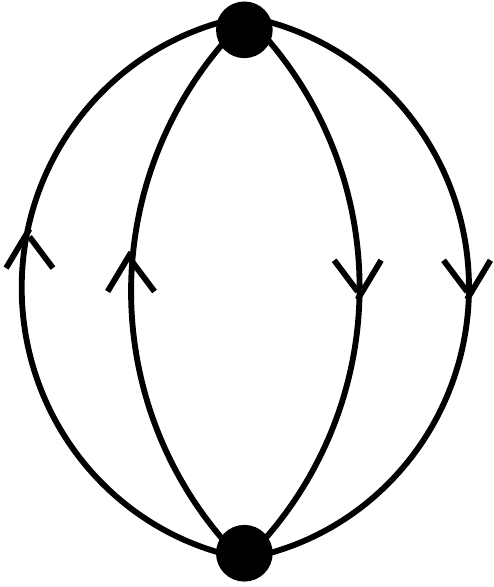}
\caption{Hugenholtz diagram for the second order contribution to the energy density.
The solid lines represent nucleons, and the black circles represent the antisymmetrized nuclear potential.}
\label{fig:3}
\end{figure}
\begin{figure}[tbp]
\centering\includegraphics[scale=0.3]{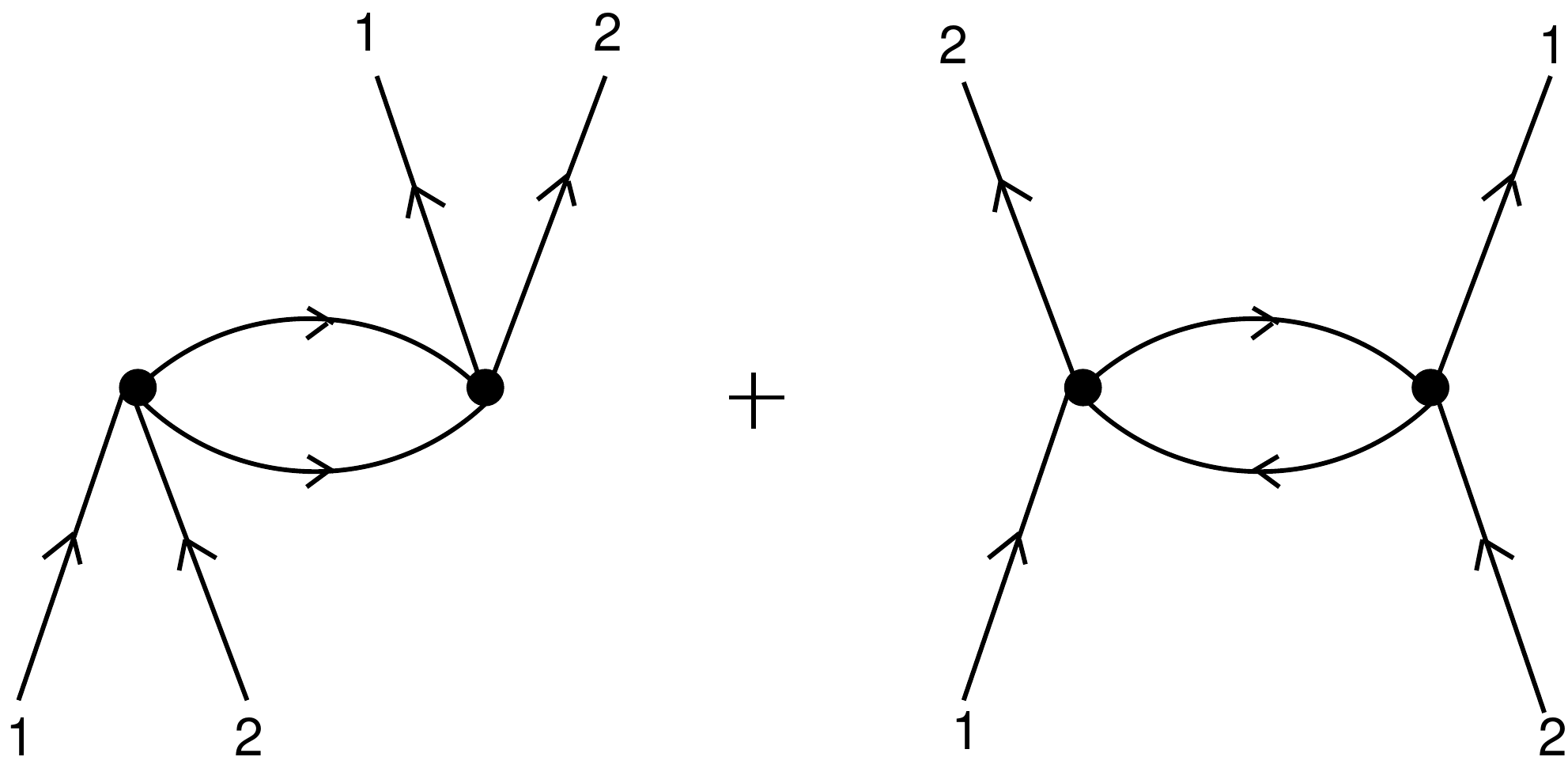}
\caption{Hugenholtz diagrams for the two-particle forward scattering amplitude. The left diagram
represents Eq.~\eqref{pp}, and the right diagram represents Eq.~\eqref{ph}. 
The solid lines represent nucleons, and the black circles represent the nuclear potential.}
\label{fig:4}
\end{figure}

For illustration of the discussions in Sec.~\ref{sec:IV}, we summarize the familiar expressions
for the energy density and the two- and three-particle forward scattering amplitudes 
in second order\footnote{In this Appendix we use a superscript $(2)$ to denote expressions
derived in  second order perturbation theory.} perturbation theory~\cite{Brown:1971zza,Negele:1988aa}.
The expression for the energy density is
\begin{align}
E^{(2)} &= - \frac{1}{4} \sum_{1234} \, \delta_{\vect{k}_1+\vect{k}_2,\vect{k}_3+\vect{k}_4}
\, \frac{|\langle 12 | \hat{v} | 34 \rangle_a|^2}{\epsilon_3 + \epsilon_4 - \epsilon_1 - \epsilon_2}
\nonumber \\
&\times n_1 \, n_2 \, \left(1-n_3\right) \, \left(1-n_4\right) \,. 
\label{e2}
\end{align}
Here we use the schematic notations explained in Sec.~\ref{sec:IV}, and 
$n_i \equiv n_{k_i} = \theta(p_F - k_i)$
is the Fermi step function. The corresponding Hugenholtz diagram~\cite{Negele:1988aa} is shown in Fig.~\ref{fig:3}. The second order contribution to the two-particle forward scattering amplitude obtained from
Eq.~\eqref{e2} is given by
\begin{align}
\lefteqn{f^{(2)}(\vect{k}_1, \vect{k}_2; \rho) = \frac{\delta^2 E^{(2)}}{\delta n_1 \, \delta n_2}}
\nonumber \\
&  =  \frac{1}{2} \sum_{34} \,  \delta_{\vect{k}_1+\vect{k}_2,\vect{k}_3+\vect{k}_4}
\, \frac{|\langle 12 | \hat{v} | 34 \rangle_a|^2}{\epsilon_3 + \epsilon_4 - \epsilon_1 - \epsilon_2} \, 
\left(n_3 + n_4 -1 \right) \label{pp} \\
 & +  \sum_{34} \,  \delta_{\vect{k}_1+\vect{k}_3,\vect{k}_2+\vect{k}_4} \, 
\, \frac{|\langle 13 | \hat{v} | 24 \rangle_a|^2}{\epsilon_2 + \epsilon_4 - \epsilon_1 - \epsilon_3} \, 
\left(n_3 - n_4 \right)\,. \label{ph} 
\end{align}
The corresponding Hugenholtz diagrams are shown in Fig.~\ref{fig:4}. The term Eq.~\eqref{pp} is represented by
the left diagram, where the bubble graph has two possible time orderings corresponding
to two particles or two holes in the intermediate state. The term Eq.~\eqref{ph} is represented by
the right diagram, where again the bubble graph has two possible time orderings, corresponding to 
forward and backward propagating particle-hole intermediate states. An average over the spin and isospin 
components of particles 1, 2 are
implicitly assumed in the above schematic notations.

By relabeling indices, it is easy to verify that 
\begin{align}
&f^{(2)}(\vect{k}_1, \vect{k}_2; \rho) = f^{(2)}(\vect{k}_1, \vect{k}_2; \rho=0) 
+ \sum_3 \, n_3 \nonumber \\
&\times \left[\sum_4 \delta_{\vect{k}_1 + \vect{k}_2, \vect{k}_ 3 + \vect{k}_4} \,
\frac{|\langle 12| \, \hat{v} \, | 34 \rangle_a|^2}{ \epsilon_3 + \epsilon_4 - \epsilon_1 - \epsilon_2}  
+ \left( 2 \leftrightarrow 3 \right) +  \left( 1 \leftrightarrow 3 \right)\right] \,. 
\label{new}
\end{align}
From this expression it is clear that the second order contribution to the three-particle
forward scattering amplitude:
\begin{align}
h^{(2)}(\vect{k}_1, \vect{k}_2, \vect{k}_3; \rho) = 
\frac{\delta f^{(2)}(\vect{k}_1, \vect{k}_2; \rho)}{\delta n_3},
\nonumber
\end{align}
is given by Eq.~\eqref{expr}. The only non-trivial point to note
is how the principal value emerges. To see this, one has to go back to Eq.~\eqref{e2} and add
the infinitesimal $- i \delta$ in the denominator, although this is irrelevant for the energy density
and the two-particle scattering amplitude, which are automatically real. Going through the 
derivation, one finds Eq.~\eqref{expr}, but with an additional contribution
\begin{align}
&\sum_4 \delta_{\vect{k}_1 + \vect{k}_2, \vect{k}_ 3 + \vect{k}_4} \,
|\langle 12| \hat{v} | 34 \rangle_a|^2 \, 
i \pi \delta\left(\epsilon_3 + \epsilon_4 - \epsilon_1 - \epsilon_2\right) (1 - 2 n_4) \no \\
&
+ \left( 2 \leftrightarrow 3 \right) +  \left( 1 \leftrightarrow 3 \right) \,. 
\label{im}
\end{align}
Because $k_i=p_F$ for $i=1,2,3$, the energy conserving delta function in Eq.~\eqref{im} 
requires that also $k_4 = p_F$.
In order to avoid such an unphysical imaginary part, one therefore has to define the step function
$\theta(x)$ so that 
$\theta(x=0)=\frac{1}{2}$, which is also suggested by the finite temperature form of $n_4$, 
taking the limit $\epsilon_4 \rightarrow \epsilon_F$ before the limit $T \rightarrow 0$.     

We note that Eq.~\eqref{new}, which reads symbolically as
$f(1,2) = f_{\rm vac}(1,2) + \sum_3 h(1,2,3) \, n_3$, does not hold in the general
case, only its differential form is general. 
This indicates that some care should be taken in a direct comparison of magnitudes of the 
dimensionless two-body [Eq.~\eqref{dlf}] and three-body [Eq.~\eqref{dlh}] parameters.

\begin{figure}
\centering\includegraphics[scale=0.4]{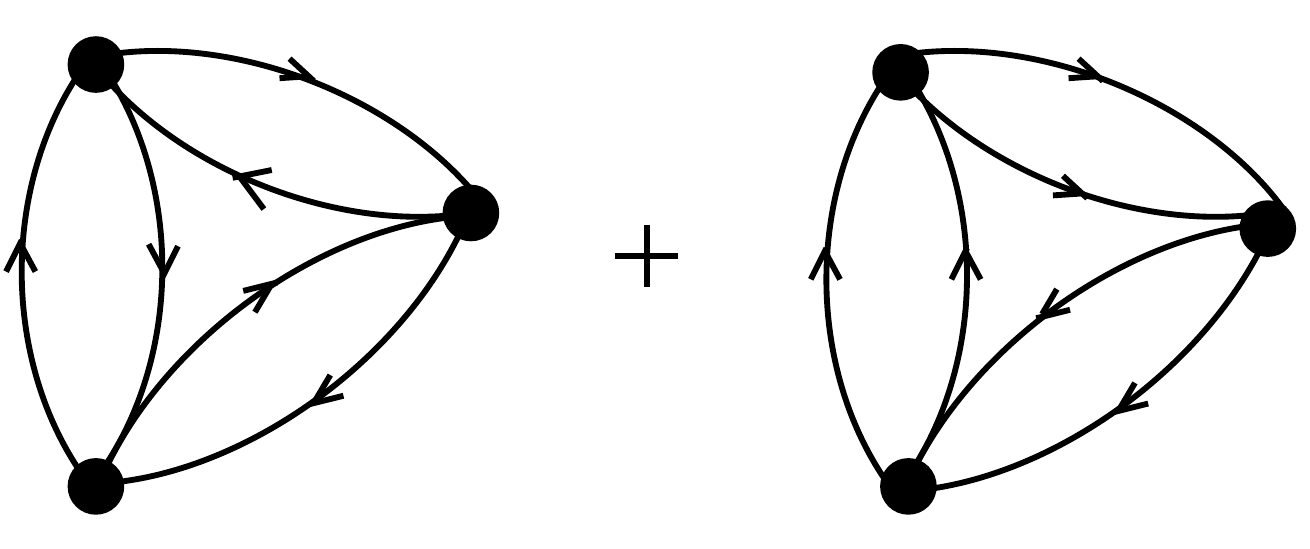} \\[1.0em]
\centering\includegraphics[scale=0.4]{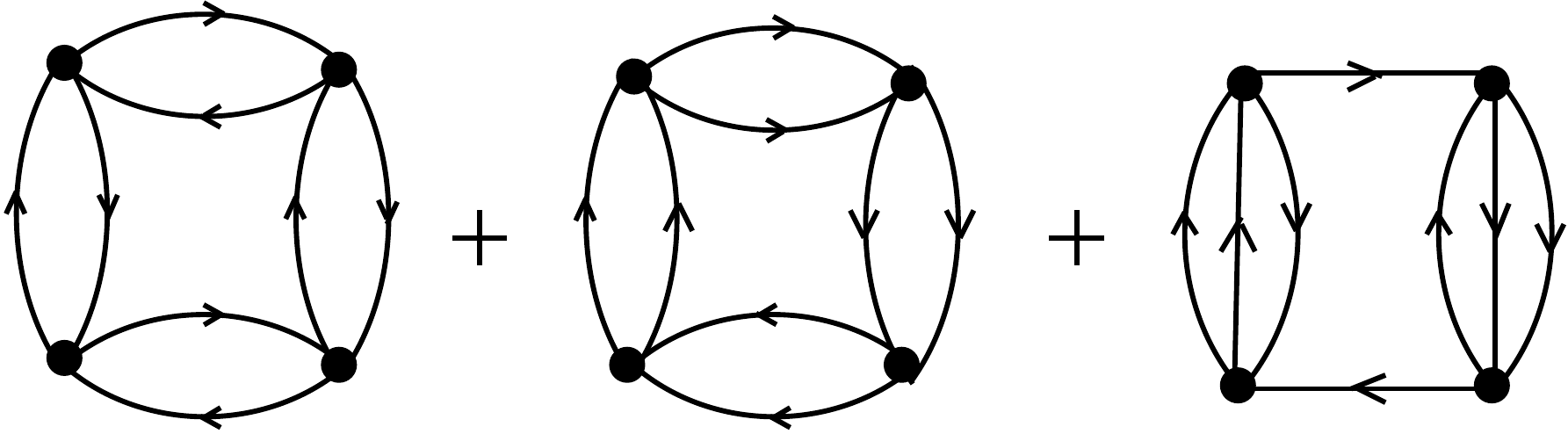}
\caption{{\it Top panel:} Hugenholtz diagram for the third order contribution to the energy density. {\it Bottom panel:} Hugenholtz diagram for the fourth order contribution to the energy density.
In both cases the solid lines represent nucleons, and the black circles represent the antisymmetrized nuclear potential.}
\label{fig:5}
\end{figure}

The discussions given above may be extended to higher orders in perturbation theory, 
and we just give the relevant Hugenholtz diagrams for the energy density in third and fourth order
in Fig.~\ref{fig:5}~\cite{Hammer:2000xg,Steele:2000qt,Bogner:2009bt}. One may use those figures to follow the line of arguments in Sec.~\ref{sec:IV}, leading from the second order
result, Eq.~\eqref{expr}, to its extension given by Eq.~\eqref{expr1}. 

%===============================================================================
%===============================================================================
   
%===============================================================================
%===============================================================================
\section{ANGULAR INTEGRALS\label{app:C}}
If we chose a coordinate system where $\vect{k}_1$ points along the $z$ axis, the energy denominator 
of the first term in Eq.~\eqref{int1} can be expressed as
\begin{align}
\lefteqn{\frac{1}{\epsilon_3 + \epsilon_4 - \epsilon_1 - \epsilon_2} \equiv
\frac{1}{ \epsilon_{\vect{k}_3} + \epsilon_{\vect{k}_1+\vect{k}_2-\vect{k}_3}-\epsilon_{\vect{k}_1} -\epsilon_{\vect{k}_2}}}
\nonumber \\
&\hs*{5mm}
= \frac{M}{p_F^2} \frac{1} {(1+z_2)(1-z_3) - \sqrt{(1-z_2^2)(1-z_3^2)} \, \cos \phi} \,.
\end{align}
Here $z_2,\,z_3$ are the polar angles of $\vect{k}_2, \vect{k}_3$, and $\phi$ is the relative
azimuthal angle between $\vect{k}_2$ and $\vect{k}_3$. The principal value integration over
$\phi$ can be performed by using the formula
\begin{align}
P \, \int_0^{2\pi} \frac{{\rm d}\phi}{ a + b \cos \phi} = \frac{2\pi}{\sqrt{a^2-b^2}} 
\, \theta(a^2-b^2),
\end{align}
where $a>0$. This gives
\begin{align}
\int \frac{{\rm d} \Omega_{2}}{4 \pi} \,\int \frac{{\rm d} \Omega_{3}}{4 \pi} \,
P \, \frac{1}{ \epsilon_3 + \epsilon_4 -\epsilon_1  -\epsilon_2}
&= \frac{M}{p_F^2} \, \frac{\sqrt{2}}{8} \, I_0 \,, \label{j0} \\
\int \frac{{\rm d} \Omega_{2}}{4 \pi} \,\int \frac{{\rm d} \Omega_{3}}{4 \pi} \,
P \, \frac{\vect{\hat k}_1 \cdot \vect{\hat k}_2}{ \epsilon_3 + 
\epsilon_4  -\epsilon_1  -\epsilon_2  }
&= \frac{M}{p_F^2} \,\frac{\sqrt{2}}{8} \, I_1 \,,  \label{j1}
\end{align}
where the integrals $I_0$ and $I_1$ over the polar angles are evaluated as
\begin{align}
I_0 &= \int_{-1}^1 {\rm d} z_2 \, \int_{-1}^{z_2} {\rm d} z_3 \,
\frac{1}{\sqrt{(1+z_2) (1-z_3)(z_2-z_3)}} \nonumber \\
&= 4 \sqrt{2} \, \ln 2 \,,  
\label{val1}
\end{align}
\begin{align}
I_1 &= \int_{-1}^1 z_2 \, {\rm d} z_2 \, \int_{-1}^{z_2} {\rm d} z_3 \,
\frac{1}{\sqrt{(1+z_2) (1-z_3)(z_2-z_3)}} \nonumber \\
&= \frac{4}{3} \sqrt{2} \, (1 - \ln 2) \,. 
\label{val2}
\end{align}
The remaining two terms in Eq.~\eqref{int1} can be obtained from the above integrals by a redefinition
of integration variables, and the results are 
\begin{align}
&\int \frac{{\rm d} \Omega_{2}}{4 \pi} \,\int \frac{{\rm d} \Omega_{3}}{4 \pi}
\left(\frac{P}{ \epsilon_3 + \epsilon_4 - \epsilon_1  -\epsilon_2}
+ (1 \leftrightarrow 3) + (2 \leftrightarrow 3) \right) \nonumber \\
&= \frac{M}{p_F^2} \frac{3 \sqrt{2}}{8} I_0  \,,  
\end{align}
\begin{align}
&3 \, \int \frac{{\rm d} \Omega_{2}}{4 \pi} \,\int \frac{{\rm d} \Omega_{3}}{4 \pi} \, 
\left(\vect{\hat k}_1 \cdot \vect{\hat k}_2\right) \nonumber \\
&\hs*{20mm}
\times \left( \frac{P}{\epsilon_3 + \epsilon_4  -\epsilon_1 -\epsilon_2} 
+ (1 \leftrightarrow 3) + (2 \leftrightarrow 3) \right) \nonumber \\ 
&=  - \frac{M}{p_F^2} \frac{3 \sqrt{2}}{8} I_1 \,,
\end{align}
which gives Eqs.~\eqref{int1} and \eqref{int2} of the main text.

%===============================================================================
%===============================================================================
%\bibliographystyle{apsrev4-1}
%\bibliography{bib}


\end{document}